# Evaluating Digital Sine wave Generator Using Analog Metrics

HoseinAli Jafari Abeshoori
Electrical engineering school
Iran university of science and technology
Tehran, Iran
hoseinalijafari@elec.iust.ac.ir

Seyed Javad Azhari
Electrical engineering school
Iran university of science and technology
Tehran, Iran
azhari@iust.ac.ir

*Abstract*— This work aims to relate comparison metrics for both Direct Digital Synthesizers (DDS) and their analog counterparts. The proposed metrics are Total Harmonic Distortion (THD) and maximum absolute error. Error is theoretically formulated into closed forms for known systematic parameters of DDS, sample rate, and bit counts. By the use of Matlab scripting, the system model is simulated for a wide range of parameters sweep.

## I. Introduction

With the rapid growth of digital systems and taking up a major portion of electronics and computer technology, some analog systems are being replaced by their digital counterparts. Analog circuits are being developed for more configurability[1] and the future of Field Programmable Analog Arrays (FPAA) is promising. So in very low power and high speed with limited accuracy applications[2], [3], digital systems substitution should be done with care. Applications with heavy computation load[4] or low output speed with no concern about consuming power are high priority nominees for substitution e.g. controllers[5] and biomedical devices[6].

It is very interesting to have common evaluating merits between digital and analog systems. Two parameters of this kind are Total Harmonic Distortion (THD) and error which is describing the difference between a target or reference signal and produced output in frequency and time domain[7]–[9].

Direct Digital Synthesizers (DDS) usually have a Low Pass Filter (LPF) to suppress very high-frequency components in the produced output signal[10]. There is the same block in analog signal generators to remain most of the signal energy in the desired frequency range. Because of the sameness in this stage between analog and digital synthesizers and added complexity for theoretical analyzing, systems mentioned here have no LPF at their output stages. This leads to more simplification and specialized analytical study. Also, the internal structure of DDS is not mentioned and it is just mentioned as a black computational box with a known Digital-to-Analog Converter (DAC) at the output node.

In the most basic form, the monotonic pure sine wave with fixed amplitude, phase, and frequency is chosen as the target output signal.

Section II will introduce the target signal in a closed formed and try to formulate the output of the DDS in three different conditions. By use of achieved formulas, their distance from the target signal is calculated and theoretically analyzed. In section III simulation results for different value of the system, parameters are presented and their accordance with the theoretical analysis described.

## II. Methode

The ideal expected output signal from DDS and its analog counterpart is a pure sine wave called target expressed in (1) with unity amplitude, real frequency, and zero phase:

$$f_0(t) = \sin(2\pi . f . t) \qquad (1)$$

In (1) f and t represent the frequency and time in Hertz and Seconds respectively. Error and harmonics distortion of other synthesized signals are calculated in comparison with the amplitude and Fourier analysis of the target function.

### A. Quantization (domain discretization) effect

The limited number of amplitude levels in digital-to-analog conversion is usually mentioned as quantization of amplitude. The values between these available levels are transferred to the nearest level. Choosing the nearest available level for a value could be performed at least in 3 different manners with floor, round, or ceiling functions they are formulated in (2), (3), and (4).

$$f_{Q,floor}(t,B) = \lfloor \sin(2\pi f t) \times 2^{B-1} \rfloor / 2^{B-1} \qquad (2)$$

$$f_{Q,round}(t,B) = \frac{\lceil \sin(2\pi f t) \times 2^{B-1} + \tfrac{1}{2} \rceil}{2^{B-1}} \qquad (3)$$

$$f_{Q,ceiling}(t,B) = \lceil \sin(2\pi f t) \times 2^{B-1} \rceil / 2^{B-1} \qquad (4)$$

Where B is the number of input bits provided by the Digital to analog converter. Consequently, the maximum value of the absolute error is just under (5) that is inversely proportional to the number of bits.

$$|f_Q - f_0| < \tfrac{1}{2^{B-1}} \qquad (5)$$

Since the total number of available analog amplitude levels in a DAC is $2^{No.BITs}$, for sine wave with both positive and negative domain, full-scale range (F.S.R.) is exactly 2 as calculated in (6) but the error in percent is usually intended by a division of peak absolute error by one.

$$F.S.R. = \sin\left(\tfrac{\pi}{2}\right) - \sin\left(-\tfrac{\pi}{2}\right) \qquad (6)$$

### B. Time discretizing (Sampling) effect

The discretization of the time factor in the digital synthesizer circuit is theoretically modeled (7) by use of floor function and including the effect of the time gap between producing of new values at the output node. The time gap Δt is mentioned to be constant for each output to become ready.

$$f_C = \sin\left(2\pi f \times \left\lfloor \frac{t}{\Delta t} \right\rfloor \times \Delta t\right) \quad (7)$$

The function $f_C$ has modeled a DAC with no loss of precision because of the limited number of available levels in output. With a given time gap, the clock pulse frequency of the digital synthesizer cannot be less than $f_{CLK}$ (8).

$$f_{CLK} = \frac{1}{\Delta t} \quad (8)$$

The absolute maximum error of the $f_C$ is difficult to be calculated into a closed-form because it has fractures for every $\Delta t$. since $f_C$ is not completely continuous, it is not fully differentiable at all its domain. To find the maximum of its absolute error, a dominant function is used.

The only difference between $f_C$ and $f_0$ is their time variable. The most distance between these time variables occurs just before the endpoint of each $\Delta t$ time gap. Now by use of (9):

$$t - \Delta t \leq \left\lfloor \frac{t}{\Delta t} \right\rfloor \Delta t \leq t \quad (9)$$

The farthest distance is less than (10):

$$\|f_C - f_0\| < |\sin 2\pi f t - \sin 2\pi f(t - \Delta t)| \leq 2 \quad (10)$$

Extremums of inequality (10) happens just before balance points of inequality (9), $t = K\Delta t$, which led to (11):

$$\|f_C - f_0\| < |\sin 2\pi f K \Delta t - \sin 2\pi f(K-1)\Delta t| \quad (11)$$

Alternative form of (11) expressed in (12) where $\omega$ is the constant factor $2K\pi f$:

$$\|f_C - f_0\| < \left|\sin(\omega \Delta t) - \sin\left(\omega \Delta t - \frac{\omega}{K}\Delta t\right)\right| \quad (12)$$

$\Delta t$ in (12) inserts a phase shift while the peak distance between two sine functions is just related to that. The greatest phase shift is calculated in (13) with taking K=1:

$$\varphi_{max} = 2\pi f . \Delta t \quad (13)$$

For very large values of time gap when $2f.\Delta t=1$, the phase difference becomes 180° and absolute error can reach up to 2. This happens when $\Delta t$ is half the target sine wave period. For small enough values of $\Delta t$, peak error can be estimated to (14) which is in direct proportion with the time gap:

$$error_{max} < \sin 2\pi f \Delta t \quad \text{when: } \Delta t \ll \frac{1}{f} \quad (14)$$

### C. Digitizing effect

A real digital synthesizer has neither an infinite number of output levels nor infinite clock frequency. In other words, a limited number of output levels and the existence of a stop time gap between output changes led to fractures in output on the time and domain axis. Including both discretization effects is necessary for investigating more practical conditions that are formulated in (15).

$$f_D(t, \Delta t, B) = \frac{\left\lfloor \sin\left(2\pi f . \left\lfloor \frac{t}{\Delta t} \right\rfloor . \Delta t\right) \times 2^{B-1} \right\rfloor}{2^{B-1}} \quad (15)$$

With the same manner for (5) and (14), the maximum error is limited to (16):

$$|f_D - f_0| \leq \frac{1 + \left\lfloor 2^{B-1} \sin(2\pi f \Delta t) \right\rfloor}{2^{B-1}} \quad (16)$$

## III. SIMULATION RESULTS

### A. Domain quantized

In a very fast system with high enough clock pulse speed and a negligible time gap for updating output value, signal specifications can be estimated by just mentioning the quantization effect on domain value. The results of the theoretical analysis for the quantization effect with a different number of bits are plotted in fig. 1 and fig. 2.

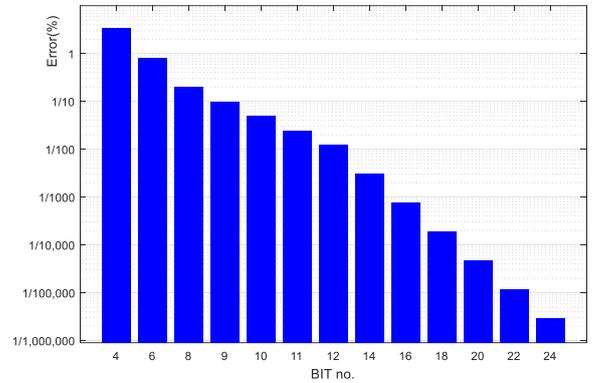

Fig. 1. Maximum absolute error in percent for some usual bit counts in logarithmic scale

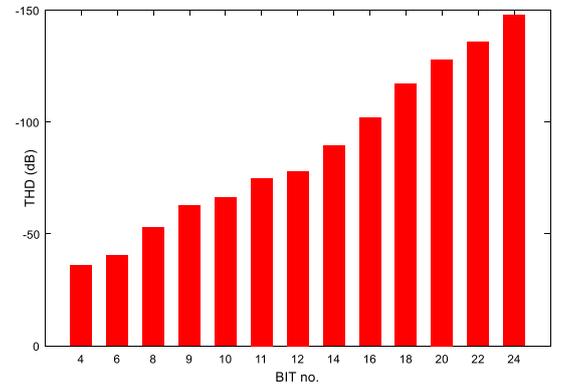

Fig. 2. THD in negative decibel scale vs. number of bits

Both error percent and total harmonic distortion (THD) diminish rapidly in a logarithmic manner by adding each extra single bit. Results for only more usual bit counts are depicted for better obviousness.

### B. Time discretized

As the bits number of a digital synthesizer approaches infinity, its domain levels get much closed and lose its negative effects on signal specifications. So the only dominant limitation remained is discontinuity in the time domain. The effect of this circumstance is investigated and sketched in fig. 3 and fig. 4.

The parameter on the horizontal axis is frequency multiplier which is the ratio of the Sine wave period to the time gap between output updates. Frequency multiplier expresses how many times the digital synthesizer is faster than its target sine wave.

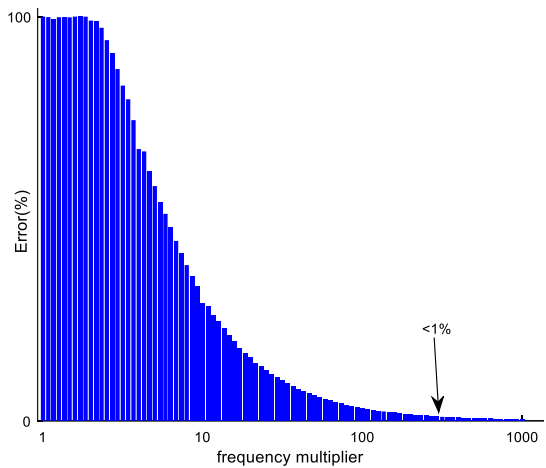

Fig. 3. Maximum absolute error in a linear scale on the vertical axis vs. a logarithmic range of frequency multiplier

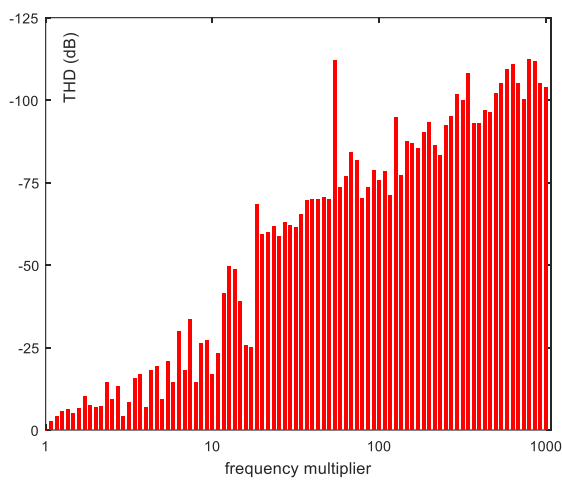

Fig. 4. Harmonic distortion is scaled in negative decibel for 30 values of frequency multiplier in each decade.

As is expected from (14), speeding up the system by making the time gaps smaller, slightly decreases the error and harmonic distortion. Notice that the horizontal axis depicted in the logarithmic scale to include larger speed-up ratios and so on the relation between THD and frequency multiplier is almost linear but for error, results are somewhat better. It is because has a reciprocal ratio of time gap Δt.

*C. Digitized signal*

In most real conditions, none of the domain quantization or time discontinuity are negligible and affect the specifications of the output simultaneously. To investigate their effect on the error and harmonic distortion, both parameters are swept and the result for each simulation depicted with colored contours of fig. 5 and fig. 6.

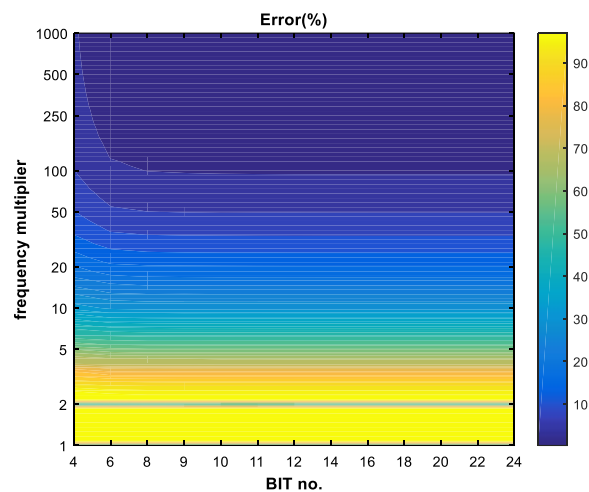

Fig. 5. The color range of yellow to blue represents the maximum absolute error of one 100 percent to zero.

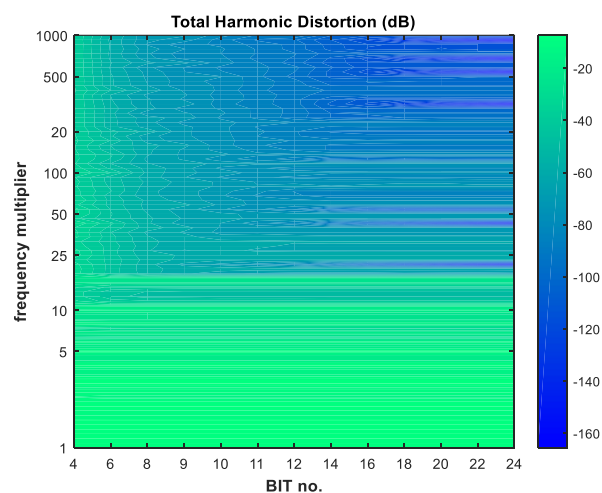

Fig. 6. lighter points show produced signal with more distorted harmonic components energy

From colored contours of fig. 5 and 6 it is obvious that for fewer bit counts, frequency multiplier is more affective and vice versa.

## IV. Conclusion

To compare an analog signal generator and its digital counterpart, the common specifications are error and harmonic distortion. This work related these properties to bits number and clock pulse speed of a general direct digital synthesizer. For future works it is recommended to investigate digital synthesizing algorithms and error metrics for a comprehensive comparison. Theoretical analysis of relation between harmonic distortion and the parameters of the digital synthesizer is absent here and needs to be explored.